\documentclass{arxiv-icarus}

\begin{document}

\begin{frontmatter}
\title{The \SI{3.4}{\um} absorption of the Titan's stratosphere: contribution of ethane, propane, butane and complex hydrogenated organics}

\author[GSMA]{Thibauld Cours}\contact{thibaud.cours@univ-reims.fr}
\author[GSMA]{Daniel Cordier}
\author[JPL,GSMA]{Beno\^{i}t Seignovert}
\author[Nature]{Luca Maltagliati}
\author[IPR]{Ludovic Biennier}

\address[GSMA]{Universit\'{e} de Reims Champagne Ardenne, CNRS, GSMA UMR 7331, 51097 Reims, France}
\address[JPL]{Jet Propulsion Laboratory, California Institute of Technology, Pasadena, CA 91109, USA}
\address[Nature]{Nature Astronomy, Springer Nature, 4 Crinan Street, N1 9XW London, UK}
\address[IPR]{Institut de Physique de Rennes, UMR CNRS 6251, 35042 Rennes, France}

\begin{abstract}
The complex organic chemistry harbored by the atmosphere of Titan has been investigated in depth by Cassini observations.
Among them, a series of solar occultations performed by the VIMS instrument throughout the 13 years of Cassini revealed a strong absorption centered at \SI{3.4}{\um}.
Several molecules present in Titan's atmosphere create spectral features in that wavelength region, but their individual contributions are difficult to disentangle.
In this work, we quantify the contribution of the various molecular species to the \SI{3.4}{\um} band using a radiative transfer model.
Ethane and propane are a significant component of the band but they are not enough to fit the shape perfectly, then we need something else.
Polycyclic Aromatic Hydrocarbons (PAHs) and more complex polyaromatic hydrocarbons like Hydrogenated Amorphous Carbons (HACs) are the most plausible candidates because they are rich in \ce{C-H} bonds.
PAHs signature have already been detected above \SI{900}{km}, and they are recognized as aerosols particles precursors. High similarities between individual spectra impede abundances determinations.
\end{abstract}

\begin{keyword}
Titan \sep Planets and Satellites \sep Solar System
\DOI{10.1016/j.icarus.2019.113571}
\end{keyword}

\end{frontmatter}


\section{Introduction}

Titan is an extraordinary object among the planets and satellites of the solar system. Its thick atmosphere,
mainly composed of nitrogen and methane, harbors a complex photochemistry producing organic compounds that participate to the formation
of haze which produces its typical orange/brown color. Due to its unique and exotic properties, Titan's atmosphere remains a very active field of research in investigating the possible origin and main properties \citep{Johnson2016,Charnay2014a,Newman2016}.
Its composition can be modeled by complex chemical network \citep{Krasnopolsky2014,Lavvas2015}, but also by laboratory experiments
\citep{Bourgalais2016,Romanzin2016} to match the observations made by Cassini's instruments
\citep{Vinatier2015,Coustenis2016,Bellucci2009}. In \cite{Bellucci2009}, the authors raised an issue about the \ce{CH4} \SI{3.3}{\um} band.

In 2006, during Cassini's 10\textsuperscript{th} flyby of Titan (T10), \cite{Bellucci2009} observed features in \ce{CH4} \SI{3.3}{\um} band with the Visible and Infrared Mapping Spectrometer (VIMS). VIMS is an imaging spectrometer onboard the Cassini spacecraft. This instrument is composed of a visible channel (\SIrange{0.3}{1.05}{\um}) and a infrared channel (\SIrange{0.89}{5.1}{\um}).
The FWHM of the 256 infrared spectral pixels are in the range \SIrange{13}{20}{nm}. VIMS can be used in different observation modes corresponding to nadir, limb and occultation geometry. Our study use the latter mode. More details can be found in \cite{Maltagliati2015}. \cite{Bellucci2009} tentatively attributed the observed features in \ce{CH4} \SI{3.3}{\um} band to solid state organic compounds, similar to those observed in the InterStellar Medium (ISM)
\citep{Sandford1991,Pendleton2002}. However, this interpretation was far to be
firm and \cite{Bellucci2009} concluded that precise comparison of their data with laboratory spectrum needed in future work.

Interestingly, a similar feature had been observed by VIMS during a stellar occultations in Saturn's atmosphere \citep{Nicholson2006,Kim2012}. Continued by \cite{Kim2012}, these analysis shown that the optical-depth spectra exhibit a broad peak at \SIrange{3.36}{3.41}{\um} for the observations obtained at 12 pressure levels between
0.0150 and \SI{0.0018}{mbar} in the Saturnian atmosphere. \cite{Kim2012}, in particular, attributed these \SI{3.4}{\um} spectral features to the aliphatic C--H stretching bands of solid-state hydrocarbons, such as \ce{C5H12}, \ce{C6H12}, \ce{C6H14} and \ce{C7H14}. \cite{Kim2011} reached a similar conclusion for Titan in analyzing the occultation T10.

\begin{figure}[!ht]
    \includegraphics[width=.9\linewidth]{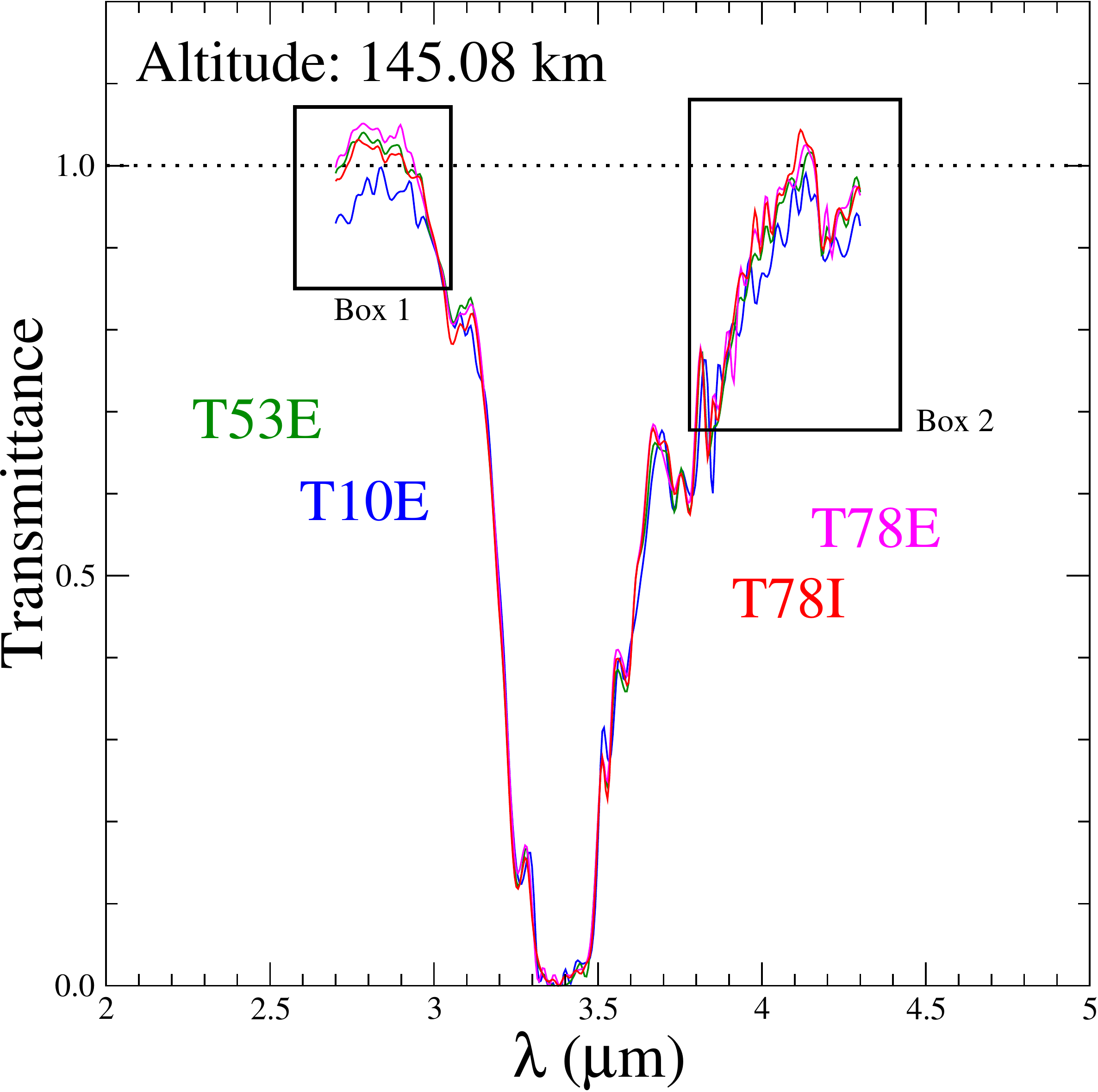}
    \caption{Transmittance curves derived from solar occultations T10E, T53E, T78E and T78I studied by \cite{Maltagliati2015}.
    Altitude interpolations, at \SI{145}{km}, have been performed to allow easy comparisons. Boxes delineate domains where data are
    significantly noisy.}
    \label{fig:estim}
\end{figure}

However, the formation of such organic microcrystals in the atmosphere of Titan is questionable due to thermodynamic arguments (see Sec.~\ref{thermod}).

More recently, \cite{Maltagliati2015} analyzed an extended set of four VIMS solar occultations, performed between
January 2006 and September 2011. In \figref{estim}, we report an example of transmission data derived from \cite{Maltagliati2015}.
Spectra have been normalized by a 3\textsuperscript{th} order polynomial and data have been interpolated at the altitude of \SI{145}{km} to allow easy comparisons. This value offers a good compromise: at lower altitudes, the stronger absorptions damp spectral features and at higher altitudes, the absorptions are weak and the spectra are more noisy due to a weaker signal.
Deviations between the four curves plotted in \figref{estim} give an idea of transmittance uncertainties, and spatial or temporal variations in Titan's atmosphere.

In their line-by-line radiative transfer model, \cite{Maltagliati2015} took into account the absorption caused by 9 molecules: \ce{CH4}, \ce{CH3D}, \ce{CO}, \ce{CO2}, \ce{C2H2}, \ce{C2H4}, \ce{C2H6}, \ce{HCN} and \ce{N2}.
In spite of the level of sophistication of their
approach, \cite{Maltagliati2015} found a clear disagreement between the observed absorption in the \SI{3.4}{\um} band and their computed spectra. It appears then that the atmosphere of Titan is a more
efficient absorber than in this model. By comparing the observed residual absorption
and the ethane cross sections measured by the Pacific Northwest National Lab \citep[PNNL - ][]{Sharpe2016}, and given the lack of \ce{C2H6}
spectral lines in major databases (HITRAN and GEISA), \cite{Maltagliati2015} interpreted the strong absorption band centered at \SI{3.4}{\um} as
the effect of ethane. These authors also tentatively attribute the narrow absorption at \SI{3.28}{\um} to the presence of Polycyclic Aromatic Hydrocarbons
(PAHs) in the stratosphere. Indeed, a few years before, \cite{Lopez-Puertas2013} identified the presence of these aromatic
molecules in Titan's upper atmosphere around \SI{650}{km} up to \SI{\sim 1300}{km}. Their results rely on the emission near
\SI{3.28}{\um} detected by VIMS.

It is well accepted that a strong absorption at \SIrange{3.2}{3.5}{\um} is related to the \ce{C-H} stretching bands, but the \ce{C-H} bonds can be present in icy hydrocarbons, in simple molecules, in more complex polyaromatic hydrocarbons or in aerosols particles. In this paper, we try to disentangle the problematic attribution of this strong
absorption around \SI{3.4}{\um}. In \figref{estim}, transmittances corresponding to box 1 and box 2 are
clearly too noisy to allow a clear analysis. In this work, we specifically focus our efforts on the highest absorption spectral range, \ie between \num{3.3} and \SI{3.5}{\um}.

In Sec. \ref{thermod} we examine, in the light of an advanced thermodynamic model, the possibility of the existence of hydrocarbon ices in the stratosphere of Titan, as proposed by \cite{Kim2011}.
In Sec. \ref{gases} and \ref{gases2} we revisit the absorption of gases in the \SI{3.4}{\um} spectral region while in Sec. \ref{pah} we discuss the possible presence of PAHs or more complex polyaromatic hydrocarbons like Hydrogenated Amorphous Carbons (HACs).
In Sec. \ref{disc} and \ref{concl}, we further discuss these issues and present our conclusions.

\section{Solid-Vapor equilibria in Titan's stratosphere}
\label{thermod}

\begin{table}[!ht]
    \caption{
        Triple points coordinates for the sample of hydrocarbon species considered by \cite{Kim2011}. The temperatures have all
        been taken in the NIST database$^{\dagger}$, for \ce{CH3CN}, \ce{C5H12} and \ce{C6H12}. The pressures are not available in
        this database, we then estimated their values by using their Antoine's equation at the triple point temperature.
    }
    \label{tab:triplepoint}
    \begin{tabular}{l l r r}
    \toprule
    Species       & Chemical       & $T_\text{triple}$ &  $P_\text{triple}$\\
                  & formula        & (K)              &  (bar)\\
    \midrule
    Ethane        & \ce{C2H6}      & 91.0             &  \num{1.1e-5}\\
    Methane       & \ce{CH4}       & 90.7             &  \num{1.2e-1}\\
    Methylcyanide & \ce{CH3CN}     & 229.3            &  \num{2.2e-3}$^*$\\
    Pentane       & \ce{C5H12}     & 143.5            &  \num{3.9e-7}$^*$\\
    Cyclohexane   & \ce{C6H12}     & 279.7            &  \num{5.3e-2}$^*$\\
    \hline
    \bottomrule
    \end{tabular}
    \hspace*{1em}\parbox{.9\linewidth}{\footnotesize
        $^{\dagger}$ \href{http://webbook.nist.gov/chemistry/}{webbook.nist.gov/chemistry}\\
        $^*$ Our estimation
    }
\end{table}

\begin{figure*}[!ht]
    \includegraphics[width=.75\textwidth]{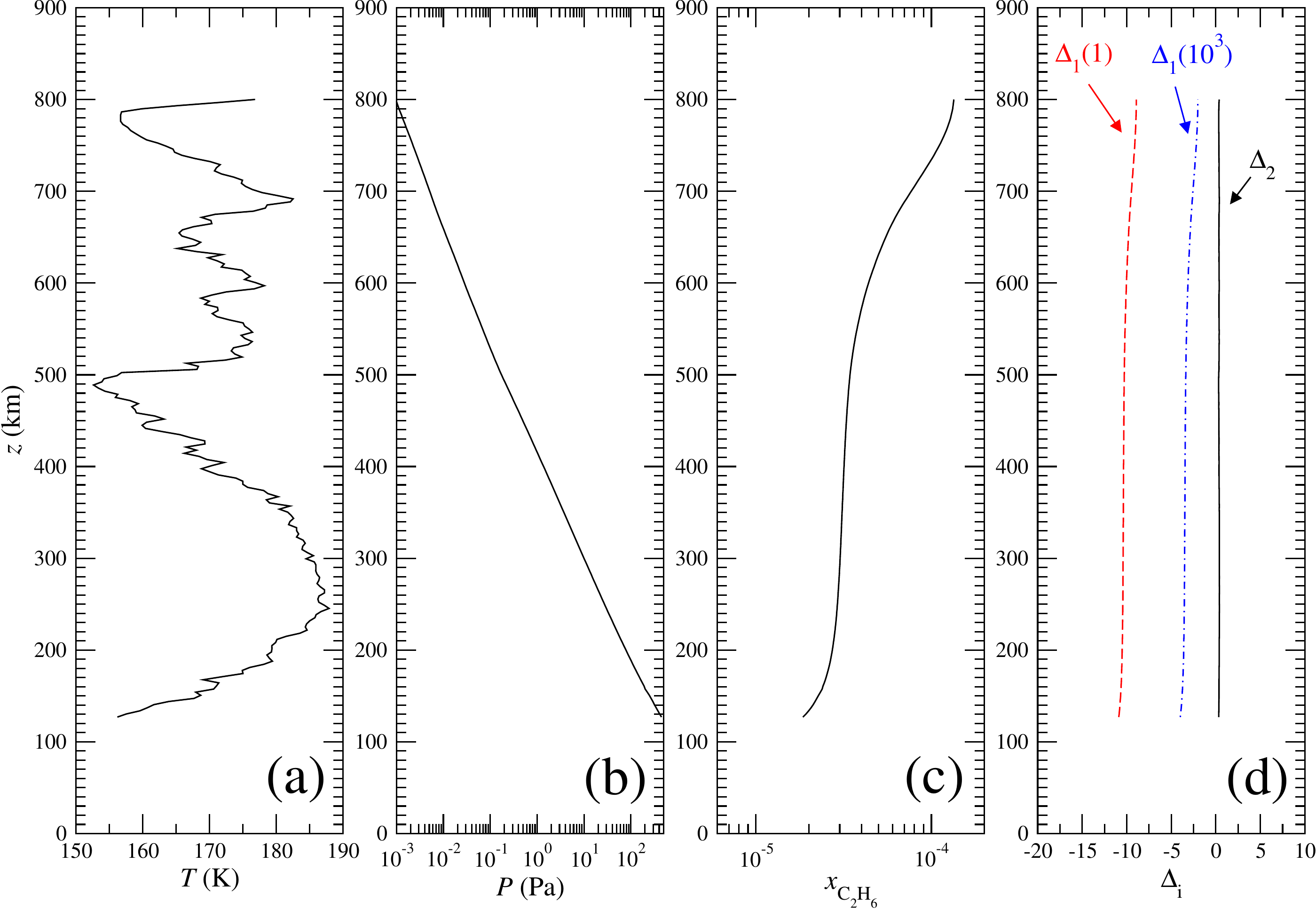}
    \caption{Panels (a) and (b) stand respectively for the temperature and pressure profiles, provided by HASI instruments, between
    the altitudes \SI{127}{km} and \SI{800}{km} in the atmosphere of Titan \citep{Fulchignoni2005}.
    Panel (c) represents the molar fraction of ethane as computed by \cite{Lavvas2008b,Lavvas2008c}.
    In panel (d) the thermodynamic quantities $\Delta_1$ and $\Delta_2$ (defined in the text by \eqref{delta_1} and
    \eqref{delta_2}) for ethane.
    We recall that $\Delta_1$ and $\Delta_2$ correspond respectively to the chemical potential, of a given species, in the gas phase and in the solid phase, the coexistence of both is reached when $\Delta_1 = \Delta_2$ (see \eqref{base}).
    $\Delta_{1}(1)$ corresponds to the abundances found by Lavvas \textit{et al.}'s while
    $\Delta_{1}(10^3)$ has been obtained by multiplying the mole fraction $x_\text{\ce{C2H6}}$ by \num{e3}.}
    \label{fig:thermoC2H6}
\end{figure*}

An explanation for the nature of Titan's \SI{3.4}{\um} absorption was put forward by \cite{Kim2011} who could reproduce the VIMS solar occultations by using hydrocarbon ices like \ce{C2H6}, \ce{CH4}, \ce{CH3CN}, \ce{C5H12} and \ce{C6H12} ices.
However, the real existence of such ices in the stratosphere of Titan needs to be discussed. Indeed, as mentioned by \cite{Kim2011}, above \SI{130}{km} (the lowest altitude explored by these authors) the temperature remains in \SIrange{160}{190}{K} interval while the pressure decreases below \SI{5e-3}{bar}. A look at \tabref{triplepoint}, in which we have gathered the triple points coordinates of involved species, shows that at least \ce{C2H6}, \ce{CH4} and \ce{C5H12} should not be in solid form at these relatively high temperatures.

In order to investigate more deeply the existence of these ices, we introduce two quantities, the first is:

\begin{equation}\label{eq:delta_1}
  \Delta_{1,i} = \ln\left( \Gamma^\text{vap}_i \, y_i \right)
\end{equation}

where $y_i$ is the mole fraction of the compound $i$ at equilibrium, in the vapor, and $\Gamma^\text{vap}_i$ is the activity coefficient of the considered species. The second introduced term is written as:

\begin{equation}\label{eq:delta_2}
 \Delta_{2,i} = -\frac{\Delta H_{i,m}}{R T_{i,m}} \, \left(\frac{T_{i,m}}{T} - 1\right)
\end{equation}

The thermodynamic equilibrium between the species $i$ in the vapor, and its icy counterpart , is reached when the equation:

\begin{equation}\label{eq:base}
  \Delta_{1,i} = \Delta_{2,i}
\end{equation}

is satisfied \citep{Poling2007}. \eqref{base} correspond to a thermodynamic equilibrium between the considered organic ice $i$ and the vapor -- \eqref{base} is an equality of chemical potential.
The activity coefficient $\Gamma^\text{vap}_i$ is given by the Perturbed-Chain Statistical Associating Fluid Theory (PC-SAFT).
Originally proposed by \cite{Gross2001}, PC-SAFT is now widely employed in the chemical engineering community, due to its very good performances.
This theory has been successfully employed in several recent studies of Titan to model liquid-vapor and solid-liquid equilibria \citep{Tan2013,Luspay-Kuti2015,Tan2015,Cordier2016b,Cordier2016c}.

In \eqref{delta_1} the activity coefficient $\Gamma^\text{vap}_i$ quantifies the \emph{degree of ideality} of the considered gas mixture. When $\Gamma^\text{vap}_i \sim 1$ the system has an ideal behavior, \ie all the molecules of the same species and those of different species interact with the same intensity.
In our context, for all the molecules listed in \tabref{triplepoint}, we found the $\Gamma^\text{vap}_i$'s very close to unity, whatever the altitude. This indicates an ideal behavior of the gases, which is not a surprise at densities provided by HASI measurements \citep{Fulchignoni2005}. Practically, this means that \eqref{base} can be satisfied only if $\Delta_{2,i}$ is negative, meaning that the molar fraction $y_i$ has to be smaller than unity.
In the case of ethane, $\Delta_{2,i}$ remains slightly larger than zero (see \figref{thermoC2H6}.d), then, in the conditions of pressure and temperature in the Titan's stratosphere, according to the present model, \ce{C2H6} can never form solid particles.
For the other species, even if $\Delta_{2,i} < 0$, the measured or estimated values of their mole fractions are order of magnitude too small for \eqref{base} to be satisfied.
For instance, concerning \ce{CH3CN}, \cite[see their Fig. 10 p279]{Lara1996} reported abundances, derived from ground-based millimeter-wave observations \citep{Bezard1993}, between roughly \num{e-9} and \num{e-7} in molar fraction, while \cite{Lavvas2008b,Lavvas2008c} computed values around \num{e-8}. Our calculations show that even in the most favorable case (\ie when the abundances of \ce{CH3CN} is taken equal to \num{e-7}) the term $\Delta_{1,i}$ is more than ten orders of magnitude smaller than the typical value of $\Delta_{2,i}$ in stratospheric conditions.
As a consequence, we conclude that the compounds proposed by \cite{Kim2011} cannot exhibit solid-vapor equilibria in the stratosphere of Titan. The only possibility remaining is the presence of these icy hydrocarbon aerosols in a non-equilibrium state, but such a situation seems unlikely.

\section{Radiative transfer modeling of the \SI{3.4}{\um} absorption}
\label{gases}

\begin{figure}[!ht]
    \includegraphics[width=\linewidth]{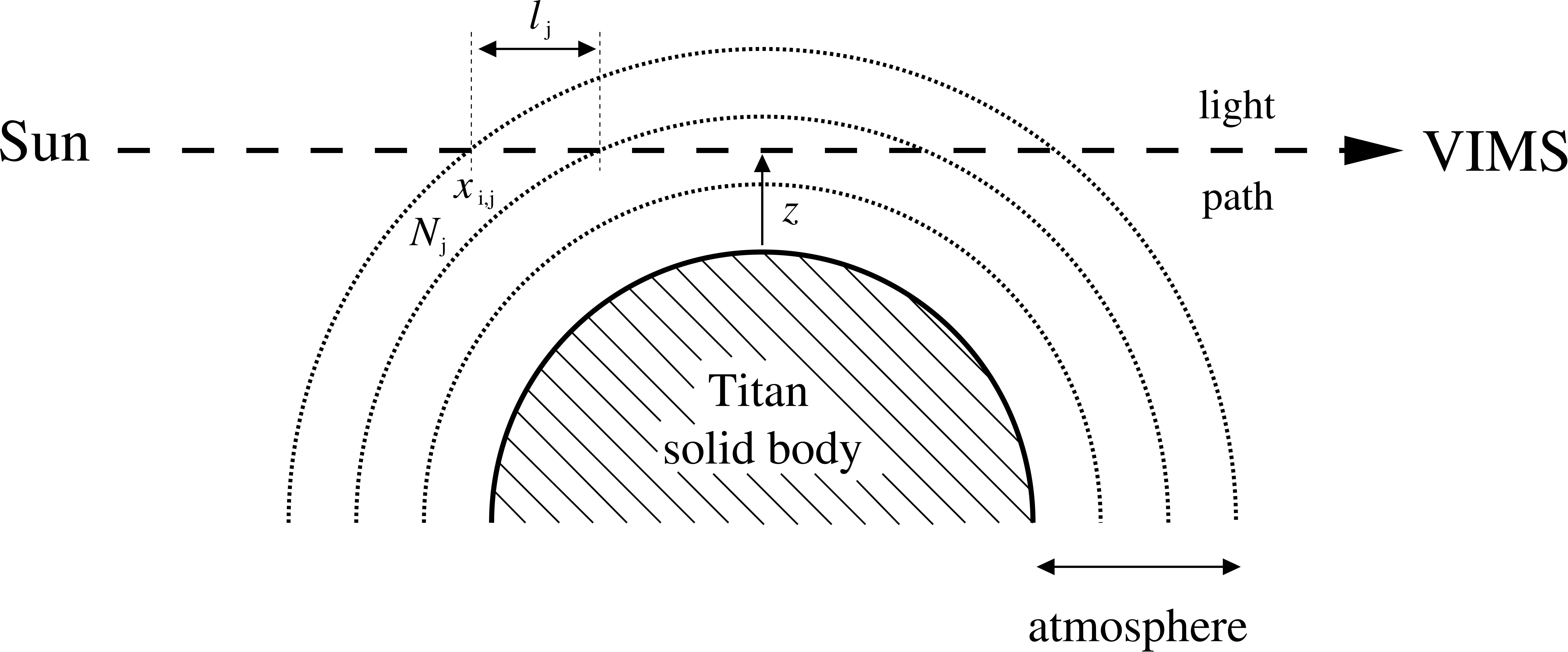}
    \caption{A schematic representation of our \emph{onion-skin} radiative transfer model, used in this paper and suitable for the
    analysis of Titan's solar occultations data published by \cite{Maltagliati2015}. In the text, the term \emph{altitude}
    refers to the parameter called $z$ in this figure. After tests with different numbers of layers, a total number $N_\text{layers}= 70$ of atmospheric layers was found sufficient to describe the atmosphere. Layers have been linearly distributed between the ground and \SI{700}{km}.}
    \label{fig:onion}
\end{figure}

In order to simulate the properties of the flux of photons that emerges from the Titan's atmosphere during a solar
occultation, we have built a simple radiative transfer model. On one hand, the structure of the atmosphere is represented by a
set of concentric spherical shells; on the other hand, the solar radiations are assumed to follow a straight optical path through
the atmosphere. The refraction is neglected in our entire approach, this approximation is relevant due to the low density probed in these explored regions. For a given altitude $z$, the transmittance $T(\lambda, z)$ of the atmosphere at the wavelength $\lambda$
is estimated using:

\begin{equation}\label{eq:transmit}
    T(\lambda,z) = \exp \left( -\sum_{i, j, k} N_j x_{i,j} \sigma_{i,k} l_j \right)
\end{equation}

where the indexes $i$, $j$ and $k$ denote respectively the chemical species, the atmospheric layers and the spectral lines.
The cross-sections are  $\sigma_{i,k}$, $N_j$ represents the total number of molecules (per units of volume) in layer $j$, while $x_{i,j}$ is the molar fraction of species $i$ in the same layer.
Finally $l_j$ stands for the distance travelled by photons in the layer $j$ (see \figref{onion}).
Doing several tests on the number of layers, we found $N_\text{layers} = 70$ to be a sufficient total number of shells, linearly distributed between the ground and a maximum altitude of \SI{700}{km}.
The pressure and temperature profiles come from HASI measurements \citep{Fulchignoni2005}.
The cross-section $\sigma_{i,k}$ (in \si{cm\squared\per molecule}) are written as:

\begin{equation}\label{eq:I_F}
    \sigma_{i,k} (\tilde{\nu}) = I_{i,k}(\tilde{\nu}_{i,k}) \, f(\tilde{\nu}-\tilde{\nu}_{i,k})
\end{equation}

where $f$ is a Voigt profile, $\tilde{\nu}$ and $\tilde{\nu}_{i,k}$ are respectively the wavenumber (in \si{\per\cm}) and the spectral line $k$ wavenumber of the chemical species $i$, $I_{i,k}$ are the intensities (in cm/molecule) of the spectral line $k$ of the chemical species $i$.
In the cases where high resolution spectral data are available, the cross-section $\sigma_{i,k}$ are computed using $k-$correlated coefficient method \citep{Arking1972,Chou1980,Fu1992}.
When only low resolution spectral data are available in the literature or when the lines are wider than the spectral resolution of VIMS (\SIrange{16}{20}{nm}), the $k-$correlated coefficient method is not necessary to compute cross-section and therefore a Gauss profile has been proved to be sufficient.

Our initial composition is based on the one published by \cite{Maltagliati2015} and include the following molecules: \ce{CH4}, \ce{CH3}D, \ce{CO}, \ce{C2H2}, \ce{C2H4}, \ce{C2H6}, \ce{H2O}, \ce{C6H6} and \ce{HCN}.
In this entire paper, we adopt this list of species as our First Guess composition (hereafter FG composition), the influence of other compounds will be made by comparing what it is obtained using this FG composition.
Nitrogen is voluntarily omitted since it is extremely poor absorbent in the domain of interest. Moreover, the collision-induced effects are negligible within the band \SIrange{3.3}{3.5}{\um}.
The \ce{CH4}, \ce{C2H2}, \ce{C2H4}, \ce{C2H6}, \ce{C6H6} and \ce{HCN} vertical mixing ratio profiles come from photochemical models developed by \cite{Krasnopolsky2014}. The abundance of deuterated methane (\ce{CH3D}) with respect to the methane has been kept constant with a \ce{CH3D}/\ce{CH4} ratio of about \num{5.3e-4} \citep{Bezard2007}.
Concerning water, we performed tests including molar fractions corresponding to the highest value given by \cite{Coustenis1998}. Finally, the abundance of \ce{CO} has been taken from \cite{Flasar2005}.

The spectral data of \ce{CO}, \ce{C2H2}, \ce{H2O} and \ce{HCN} were mainly taken in HITRAN\footnote{\href{http://hitran.org/}{hitran.org}} \citep{Rothman2013} and GEISA\footnote{\href{http://www.pole-ether.fr/geisa/}{www.pole-ether.fr/geisa}} \citep{Jacquinet-Husson2008}.
For \ce{CH4}, \ce{CH3D} and \ce{C2H4} we used the up-to-date theoretical line lists computed by \cite{Rey2016} and available at the Theoretical Reims-Tomsk Spectral database\footnote{\href{http://theorets.tsu.ru}{theorets.tsu.ru}}.

For benzene, due to the lack of data in these database, we performed \emph{ab initio} computations of its absorption frequencies
and their respective intensities. Based on the \texttt{MP2/6-311G**} level of the theory, developed by \cite{Moller1934}, we obtained frequencies, which were scaled by 0.95, according to that is the recommended in such a situation.
The only none zero intensities near the \SI{3.4}{\um} band are for the frequencies \SI{3064.0917}{\per\cm} (\SI{3.26}{\um})
and \SI{3064.0949}{\per\cm} (\SI{3.26}{\um}) (including the scaling factor of 0.95) corresponding to two \ce{C-H} asymmetric
stretching normal modes. The intensities were found quite low with values of \SI{32.63}{km/mole} for both transitions.
Considering the vertical profile of \ce{C6H6}, we find that the benzene is a minor absorber around \SI{3.26}{\um}.

\cite{Maltagliati2015} selected a set of four occultations data, T10 Egress (T10E), T53 Egress (T53E), T78 Egress (T78E) and T78 Ingress (T78I) acquired during three flybys, T10, T53 and T78 respectively.
For each occultation, the altitude ranges between \SI{\sim 50}{km} and \SI{\sim 690}{km}.
We provide, in supplementary material, four cube's name lists used in our analysis, one name list by occultation. The name lists are in Excel cvs format and can be downloaded from the pds-rings site \footnote{\href{https://pds-rings.seti.org/cassini/vims/}{pds-rings.seti.org/cassini/vims}}. For all these occultations, the retrieved transmittance curves, at a given altitude, are extremely similar \citep[see for instance Fig. 11 of][]{Maltagliati2015}. In order to facilitate comparison between our theoretical output, and observational determinations, we have chosen the Ingress occultation T78 (T78I) as a typical case. We also selected the altitude of \SI{145}{km} because it offers a good compromise between high altitudes data for which the transmittance curves are pretty flat, and low altitudes measurements presenting a global strong absorption that masks or damps spectral features.

\begin{figure}[!ht]
    \includegraphics[width=.9\linewidth]{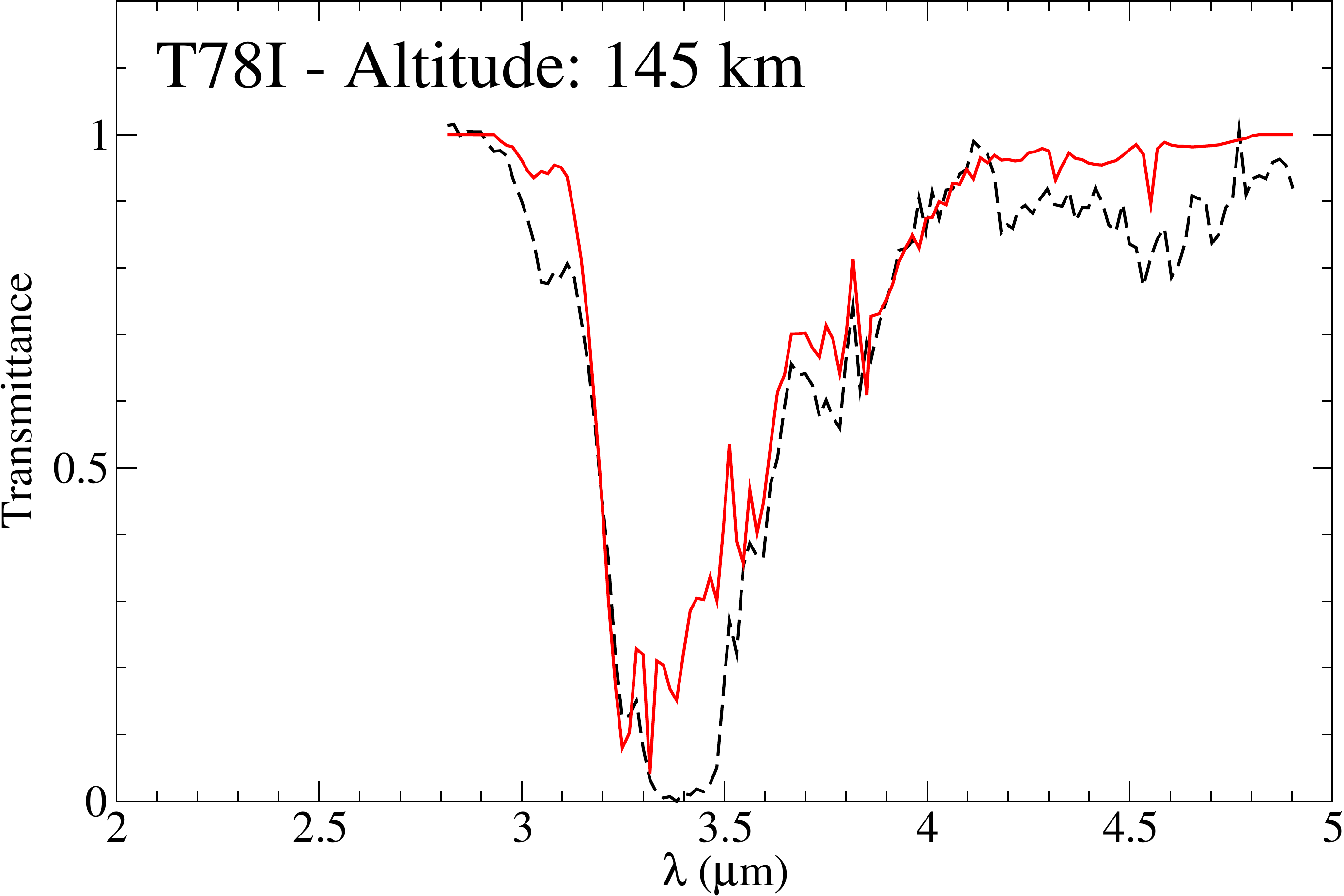}
    \caption{Comparison between observed transmittance (dashed line) acquired during occultation T78I at the
    altitude of \SI{145}{km} \citep{Maltagliati2015}, and the computed transmittance due to our FG composition:
    with only \ce{CH4}, \ce{CH3D}, \ce{CO}, \ce{C2H2}, \ce{C2H4}, \ce{C2H6}, \ce{H2O}, \ce{C6H6} and \ce{HCN}.
    For \ce{C2H6}, only spectral lines provided by HITRAN have been taken into account.}
    \label{fig:transmit145kmT78I}
\end{figure}

A first model-observation comparison can be seen in \figref{transmit145kmT78I}, the disagreement is clear, particularly
around \SI{3.4}{\um} our domain of interest. This nicely confirms the findings by \cite{Maltagliati2015}.

\section{The possible absorption of ethane, propane and butane around \SI{3.4}{\um}}
\label{gases2}

\subsection{Ethane}

\begin{figure}[!ht]
    \includegraphics[width=.95\linewidth]{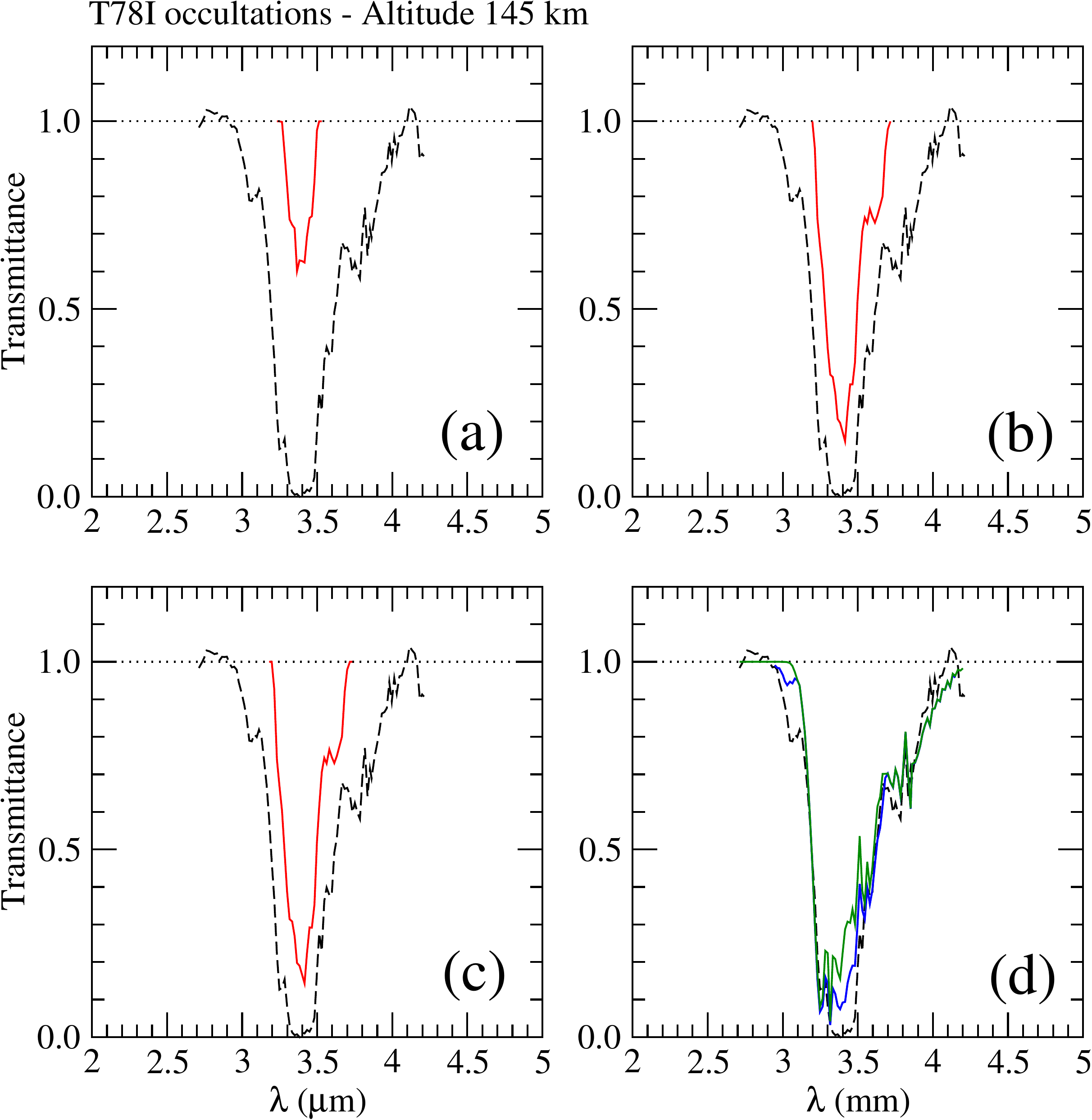}
    \caption{Comparison between simulated transmittances and VIMS for the T78I occultation data \citep{Maltagliati2015}.
    Observational data are in dashed line, the altitude is \SI{145}{km}. This figure shows the transmittance of \ce{C2H6} using:
    (a) only \cite{Pine1982} spectral lines (red), (b) only pseudo-lines lists based on the cross-section measurements \citep{Harrison2010} (red), (c) the combination of PL82 and H10 spectral data (red), (d) the combination of our FG composition model with PL82 and H10 (blue), the absorption computed with our FG composition is also plotted for comparison (green solid line).}
    \label{fig:C2H6}
\end{figure}

As already mentioned, the \ce{C-H} stretching bands produce a strong absorption at \SIrange{3.2}{3.5}{\um}, this is why any compound containing one or several \ce{C-H} bounds can potentially contribute to the observed \SI{3.4}{\um} absorption. In this context, ethane, quantitatively the main product of Titan's photochemistry \citep{Lavvas2008b,Lavvas2008c,Krasnopolsky2014}, should be the object of our first intentions. The presence of ethane in the Titan's atmosphere is firmly established by previous observations, \eg it has been detected by Cassini's instruments: UVIS \citep{Koskinen2011}, INMS \citep{Cui2009} and CIRS \citep{Vinatier2010a}. Unfortunately, in HITRAN and GEISA databases, ethane spectral lines in the band of interest are pretty scarce. Surprisingly, the absorption spectrum of the \ce{C-H} stretching region of ethane, measured by \cite{Pine1982} (hereafter PL82) at T = \SI{119}{K}, is not available in these compilations. Then, we included the \SI{\sim 3000}{lines} provided by PL82 data in our model; \num{1614} entries specify wavenumbers, intensities and the lower state energies whereas for \num{1426} other entries the lower state energy is not available. Thus, in the latter case, we have neglected the temperature corrections of the intensities. The contribution to the absorption around \SI{3.4}{\um}, of ethane alone, is plotted in \figref{C2H6}(a).

Unfortunately, \cite{Pine1982} did not include ethane PQ-branches which should have a non-negligible contribution in the domaine
of interest. To tackle the issue, we used an empirical pseudo-line list based on the cross-section measurements developed by
\cite{Harrison2010} (hereafter H10 dataset) and freely available on the web\footnote{\href{http://mark4sun.jpl.nasa.gov/pseudo.html}{mark4sun.jpl.nasa.gov/pseudo.html}}.
Compared to the absorption obtained with PL82 spectral lines, the effect of ethane is significantly enhanced by the use of this more
comprehensive list (\figref{C2H6}.b). If both, the PL82 and H10 spectral data, are simultaneously included in our model (\figref{C2H6}.c),
the actual effect of PL82 is not noticeable. Finally, if we merge PL82 and H10 spectral lines sets, with those employed for our FG composition model, the disagreement with Cassini/VIMS observations is considerably reduced (\figref{C2H6}.d).
This clearly demonstrates the prominent role of ethane, as an absorber, at wavelengths around \SI{3.4}{\um}. Nonetheless, the simulated transmittance remains significantly above the observed one. This fact suggests the presence of other absorbers, possibility which we discuss further in next paragraphs.

\subsection{Propane}

\begin{figure}[!ht]
    \includegraphics[width=\linewidth]{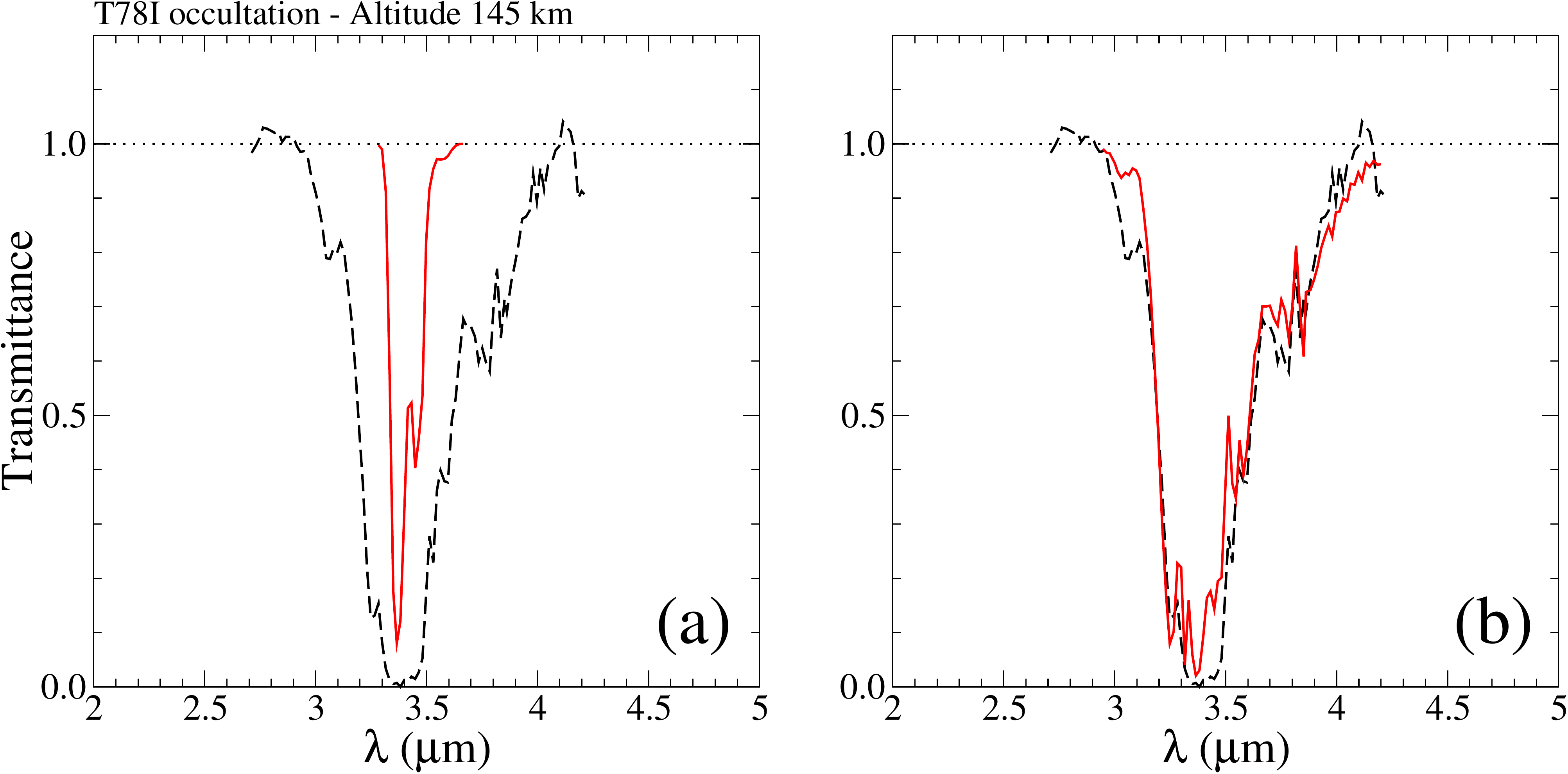}
    \caption{Occultation T78I at \SI{145}{km} (observed transmittance in dash line): (a) Absorption due to propane alone, (b) computed absorption using our FG composition complemented by propane (ethane is not taken into account).}
    \label{fig:C3H8}
\end{figure}

According to photochemical models \citep{Lavvas2008b,Lavvas2008c,Krasnopolsky2014}, propane should also be produced in
Titan's upper-atmosphere. This \ce{C3} hydrocarbon has been detected by several Cassini's instruments: \cite{Nixon2013} determined its mixing ratio using CIRS, while \cite{Cui2009} and \cite{Magee2009} retrieved abundances from INMS measurements.
Similarly to the ethane case, HITRAN and GEISA are very poor in spectral data around \SI{3.4}{\um} for this molecule. Then, we used the propane pseudo-lines list based on the cross-section measurements of \cite{Harrison2010a} (freely available on the web\footnote{\href{http://mark4sun.jpl.nasa.gov/pseudo.html}{mark4sun.jpl.nasa.gov/pseudo.html}}). Taking into account the propane abundance profile predicted by \cite{Krasnopolsky2014}, we have estimated the corresponding absorption in our domain of interest: in \figref{C3H8} we have displayed the absorption of propane alone a), the effect of this molecule is combined with that of our FG composition sample b), clearly the contribution of propane is comparable to that of ethane, even is propane is approximately one order of magnitude less abundant than ethane. In fact, due to its larger numbers of C-H, propane is an absorber roughly one order of magnitude more efficient than ethane.

\subsection{Butane}

Butane has not yet been detected, in the atmosphere of Titan. The presence of butane is predicted by photochemical models \citep{Krasnopolsky2009,Krasnopolsky2010,Krasnopolsky2014}. The quantity of butane should be lower than what it is measured and computed for propane. Available models indicate a butane mixing ratio about four order of magnitude smaller than what \cite{Krasnopolsky2014} obtained for propane.
Since no data concerning butane were found in HITRAN and GEISA databases, we used cross-sections provided by the NIST\footnote{\href{http://webbook.nist.gov}{webbook.nist.gov}}. According to this approach and taking vertical profile provided by \cite{Krasnopolsky2014}, we have checked that butane has no detectable influence on the \SI{3.4}{\um} atmospheric transmittance.

\subsection{Conclusion about ethane, propane, butane and others linear hydrocarbon}

\begin{figure}[!ht]
    \includegraphics[width=.8\linewidth]{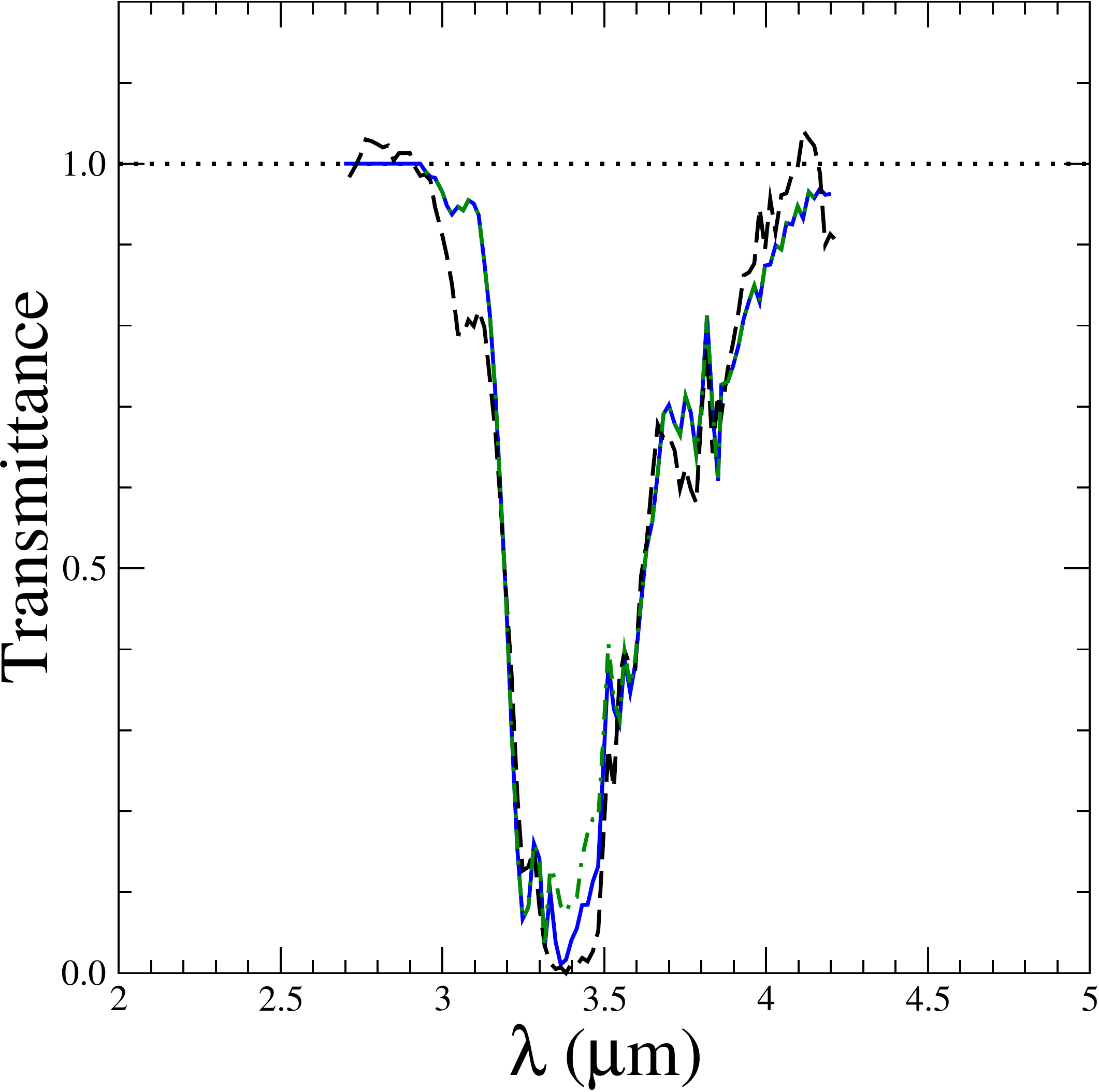}
    \caption{Dashed line: T87I observations, dot-dashed line: our computation taking into account gases of our FG
    composition and ethane (including pseudo-lines and PL82 data), solid line: the same simulation including propane pseudo-lines list based on the cross-section measurements of \cite{Harrison2010a}.}
    \label{fig:FGethaprop}
\end{figure}

We have summarized our simulations results in \figref{FGethaprop}, clearly the addition of propane to our set of considered gaseous species reduced the disagreement between theoretical output and solar occultation data. However, the situation is far from satisfactory, and there is room for other efficient absorbers around \SI{3.4}{\um}.
It is likely that other larger linear hydrocarbons are present in the atmosphere, but they are not yet retrievable from the occultation data. Moreover, their molar fractions could be very small in comparison with the molar fraction of butane. Thus, no additional linear hydrocarbon larger than butane was considered in our model.

\section{The absorption due to Polycyclic Aromatic Hydrocarbons and Hydrogenated Amorphous Carbons}
\label{pah}

As mentioned in the introduction, PAHs or more complex polyaromatic hydrocarbons like HACs \citep{Dartois2004,Dartois2005,Dartois2007}, considered as organic compounds in solid state, are detected in the ISM \citep{Sandford1991,Pendleton2002}.
For instance in \cite{Dartois2004} the features observed in the ISM spectra fitted very well with those observed in spectra of HACs produced in laboratory \citep{Dartois2004,Dartois2005}. Furthermore, PAHs are detected on the Iapetus' and Phoebe's surface \citep{Cruikshank2008}, in micro-meteorites in Antarctic \citep{Becker1997} and in the meteorite Allende \citep{Becker1997a}. Observations of comets reveal the presence of PAHs \citep{Li2009} and
in addition they were considered in Titan's upper atmosphere to explain an unidentified emission at \SI{3.3}{\um} \citep{Dinelli2013,Lopez-Puertas2013}. According to \cite{Bellucci2009} for Titan's atmosphere and \cite{Dartois2004,Dartois2005,Dartois2007,Sandford1991,Pendleton2002} for the ISM, the features observed between \num{3.38} and \SI{3.48}{\um} are due to the symmetric and asymmetric stretching of the \ce{C-H} bond in \ce{-CH_2} and \ce{-CH3} groups of the aliphatic chains. Likewise the features around \SI{3.3}{\um} are signatures of stretching of aromatic \ce{C-H} bond \citep{Bellucci2009,Dartois2004,Dartois2005,Dartois2007}. These latter signatures emphasize the presence of aromatic compounds like PAHs and the signatures at \num{3.38} and \SI{3.48}{\um} together with the signatures of aromatic \ce{C-H} stretching show the presence of complex particles containing aromatic cycles and aliphatic chains as it would expect in HACs. In \cite{Dartois2005}, possible structures of HACs compatible with the ISM spectra were simulated with a neuronal network simulation: The resulting structure shows aromatic cycles and aliphatic chains as expected.

Thus, to clarify the observed transmission in the \SIrange{3.3}{3.5}{\um} spectral region we considered PAHs and HACs compounds in our model.

\subsection{Polycyclic Aromatic Hydrocarbons}

\begin{figure}[!ht]
    \includegraphics[width=.9\linewidth]{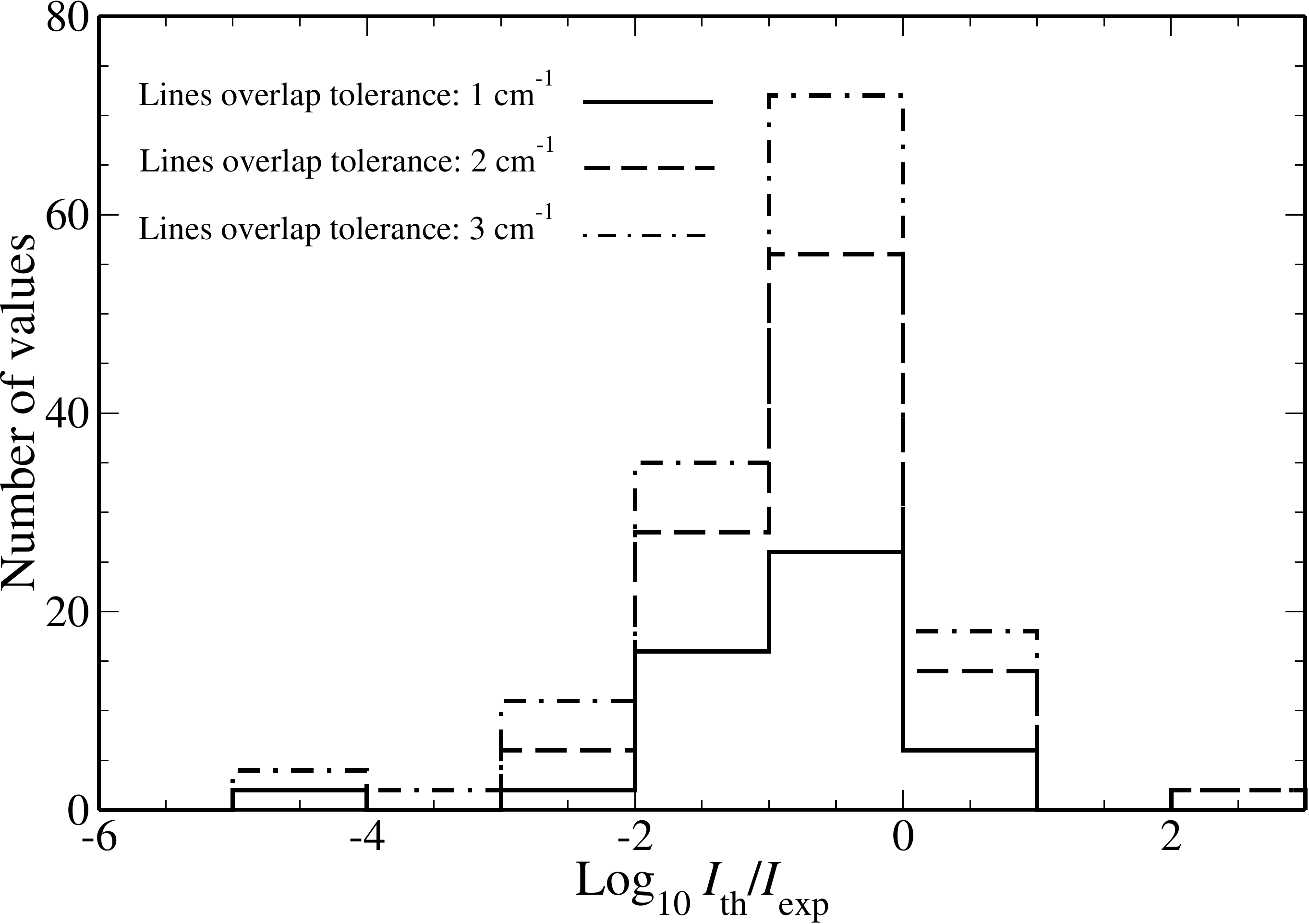}
    \caption{Histograms of the log ratios $I_\text{th} / I_\text{exp}$ for the Ames database PAHs for which theoretical and experimental
    spectral lines features determinations are available. We have considered 3 cases:
    (1) the wavelengths of theoretical and experimental match with a maximum tolerance of \SI{1}{\per\cm},
    (2) the same with \SI{2}{\per\cm} and finally \SI{3}{\per\cm}.}
    \label{fig:histoRappIthsIexp}
\end{figure}

The information on PAHs's transition intensities come from the NASA Ames PAHs IR Spectroscopic Database \footnote{\href{http://www.astrochem.org/pahdb/}{astrochem.org/pahdb}}.
In our calculations the line widths were fixed to a reasonable value of \SI{30}{\per\cm}. This value is approximately comparable to the different values used for HACs \citep{Dartois2007}.
We introduce a correction factor $\alpha_{i,k}$ for the intensities $I_{i,k}$, with the same meaning for index $i$ and $k$ than the index in \eqref{transmit}, to account for two sources of uncertainty:
(1) the temperature dependence of the intensities $I_{i,k}$,
(2) the uncertainties on the $I_{i,k}$'s themselves.
Indeed, the spectral data provided by the Ames Database are valid for a temperature of \SI{296}{K} while the actual temperature in Titan's stratosphere is substantially lower (see for instance \figref{C2H6}.a). In the spectral window of interest (\ie \SIrange{2700}{3570}{\per\cm}), for the 716 species reported in the Ames database, we counted a total of \num{11307} calculated spectral lines against only 127 coming from an experimental determination. Even if some overlaps are present, we see that the majority of available spectral data are calculated theoretically.
This rises the question of the degree of confidence that can be placed in these computed data. In this context, we have searched for theoretical and experimental spectral lines that coincide in terms of wavelength, adopting a given tolerance, respectively: \SI{1}{\per\cm}, \SI{2}{\per\cm} and \SI{3}{\per\cm}.
This way, we identified 52 lines that correspond to both theoretical and laboratory determinations with a tolerance of \SI{1}{\per\cm}, 108 when is increased to \SI{2}{\per\cm} and 144 when the tolerance is increased to \SI{3}{\per\cm}.
Consequently, we formed the log ratio $I_\text{th} / I_\text{exp}$ for each identified couple of lines, with $I_\text{th}$ the theoretical intensity and $I_\text{exp}$ the corresponding laboratory measurement.
The histograms of log ratio $I_\text{th} / I_\text{exp}$ values is plotted in \figref{histoRappIthsIexp}.
This figure shows clearly that the theoretical intensities tend to be underestimated, compared to their experimental counterparts, by a factor up to two orders of magnitude. Then, these results motivated our introduction of a factor $\alpha_{i,k}$ that represents the uncertainties that affect the spectral lines intensities available in the Ames database. In this manner, the intensity $I_{i,k}$ in \eqref{I_F} is replaced by the product $\alpha_{i,k} I_{i,k}$. Finally, if we combine \eqref{transmit} with \eqref{I_F}, the product $\beta_{i,j,k} = x_{i,j} \alpha_{i,k}$, with the same meaning for index $j$ than the index in \eqref{transmit}, appears in the expression of the transmission.
In first approach, we considered that the correction factors $\alpha_{i,k}$ are independent of the spectral lines. Therefore, $\beta_{i,j,k} = x_{i,j} \alpha_{i,k}$ becomes $\beta_{i,j} = x_{i,j} \alpha_i$.
We took 118 neutral and charged PAHs in the Ames database. We have chosen PAHs with less than 100 carbon atoms, free of \ce{Fe}, \ce{Mg}, \ce{Si} (not relevant for Titan's atmosphere) and for which the intensities are significant in the \SIrange{3.3}{3.5}{\um} band.

\subsection{Hydrogenated Amorphous Carbons}

The HACs (sometimes noted {a-C:H} or {a-C:H:N} if the HACs contain nitrogen), are considered as smallest haze particle and precursor of more complex haze particle \citep[p.304]{Lavvas2011a,Muller-Wodarg2013}.
In \cite{Dartois2004,Dartois2005,Dartois2007} no identified structure was given but just different vibration modes together with their vibrational frequencies or wavelengths.
Analyzing the spectrum of the galaxy named \texttt{IRAS 08572+3915}, \cite{Dartois2007} fitted the intensities, wavelengths and widths of the lines for these vibration modes.
Then, we use spectrum parameters given in Table 1 of \cite{Dartois2007}. We do not consider HACs with nitrogen ({a-C:H:N}) because the vibrational modes containing N are not present in \SIrange{3.3}{3.5}{\um} band \citep{Dartois2005}. As with PAHs, we also introduce a correction factor $\alpha_{i,k}$ on the intensities $I_{i,k}$, with the same meaning for index $i$ and $k$ than the index in \eqref{transmit}, taking into account for an uncertainty source: the temperature dependence of the intensities $I_{i,k}$. Indeed, the intensities coming from the study of \texttt{IRAS 08572+3915} spectrum the temperature must be different than that in Titan's stratosphere. In \cite{Dartois2007} no distinction was made between the different HACs but just between the vibration modes. Consequently, in our model for HACs we omit index $i$. Thus, our correction factor $\alpha_{i,k}$ becomes $\alpha_k$ and the intensities are $I_k$. In first approach, as with PAHs, the correction factors $\alpha_k$ are independent of the spectral lines. Consequently, we define a product $\beta_j = x_j\alpha$.

\subsection{Results}

For PAHs, $x_{i,j}$ is the molar fraction of PAHs number $i$ with respect to the \ce{C6H6} abundance profile in layer $j$.
As a first approximation, the vertical profile of this molar fraction of PAHs number $i$ vs the \ce{C6H6} (used as proxy) abundance and the vertical profile of the molar fraction of HACs were kept constant.
Thus for PAHs and HACs respectively, $\beta_{i,j}=x_{i,j} \alpha_i$ turns into $\beta_i = x_i \alpha_i$ and $\beta_j = x_j \alpha$ turns into $\beta = x \alpha$.
Considering the uncertainties on PAHs and HACs intensities, we have fitted, respectively, the product $\beta_i = x_i \alpha_i$ and not $x_i$ alone, and $\beta = x \alpha$ and not $x$, to get the best fit to occultation data.
However, the resulting values do not provide information about the actual abundances of these PAHs and HACs
because the problem is degenerated: the correction coefficient to apply remains unknown. Taking into account the log ratio $I_\text{th} / I_\text{exp}$ and according to the PAHs number $i$, we obtain as best fit for $\beta_i$, values in the range of \SIrange{4.75e-4}{3.32}{$\times$} molar fraction of \ce{C6H6}.
Concerning the HACs, the best fit for $\beta$ is found for the value \num{e-7}.

\begin{figure}[!ht]
    \includegraphics[width=.8\linewidth]{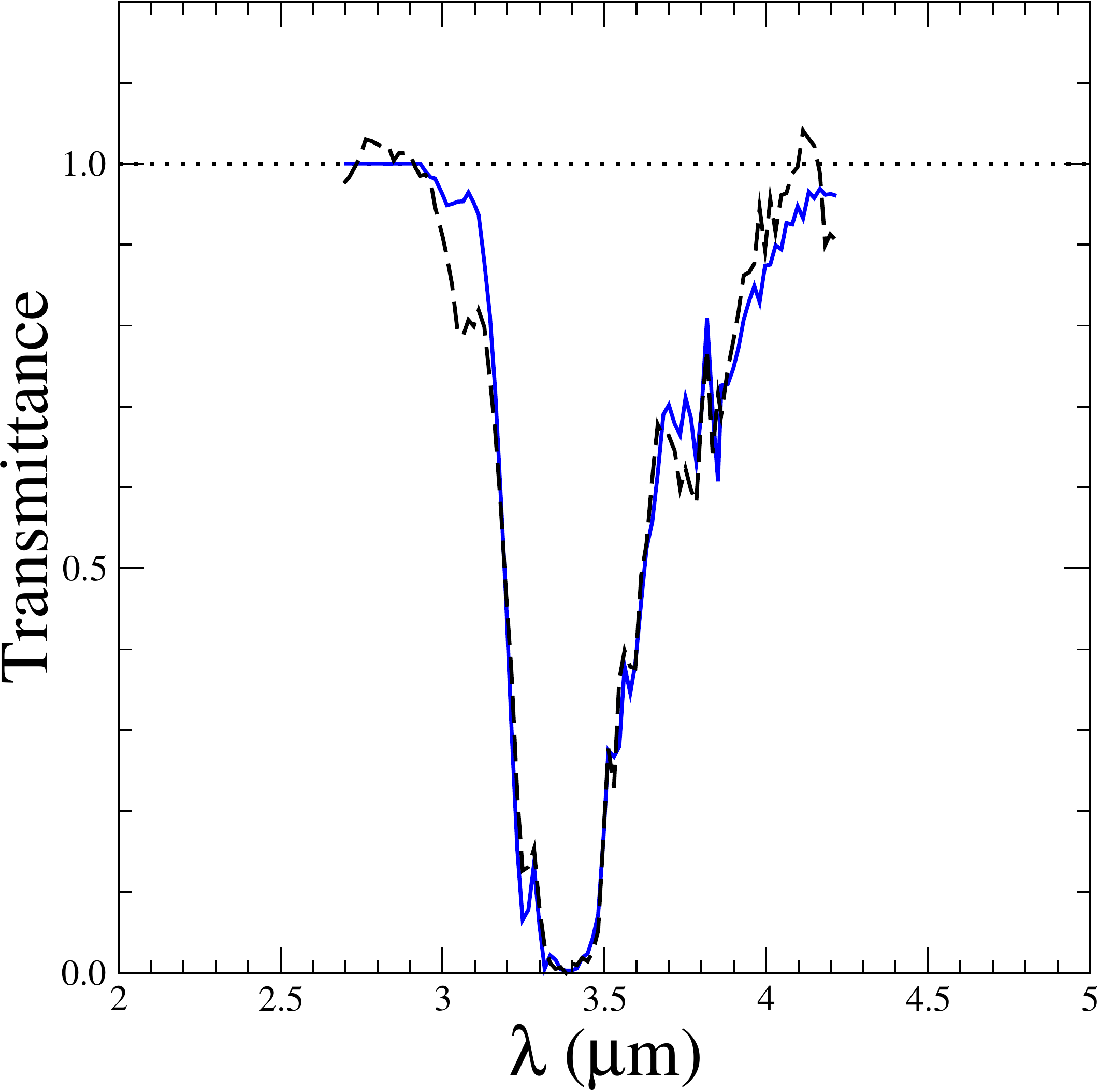}
    \caption{Computed transmittance (solid line) at \SI{145}{km} during occultation T78I (observed transmittance in dash line), the radiative transfer model takes into account all studied gases plus the best fit of 118 PAHs (from NASA Ames database) and HACs \citep[from][]{Dartois2007}.}
    \label{fig:gasesPAHHAC}
\end{figure}

\figref{gasesPAHHAC} shows that the combination of all gases, plus the 118 PAHs and HACs allow a satisfactory modelization of the observed transmittance at \SI{145}{km}.

\section{Discussion}
\label{disc}

Observations of Saturn's stratospheric auroral regions \citep{Guerlet2015} seem to underline the remarks made on the features observed in Titan's atmosphere. Indeed, spectral signatures of benzene and aerosols in the range \SIrange{680}{900}{\per\cm} and \SIrange{1360}{1440}{\per\cm} have been observed with CIRS on board Cassini, around the \ang{80}S. In particular, vibrational modes in aliphatic and aromatic hydrocarbons were observed revealing the presence of PAHs or HACs in stratosphere of Saturn \citep{Guerlet2015}. Moreover, as mentioned, VIMS observations of Procyon's occultation through Saturn's atmosphere around the \ang{55}N, reported by \cite{Nicholson2006} and discussed by \cite{Bellucci2009}, show similar features in the \SI{3.4}{\um} band to those observed in our study. Nevertheless, the observations reveal slight differences with the Titan's \SI{3.4}{\um} band spectra. But these differences could be due to a lower amount of nitrogen compounds in Saturn's atmosphere \citep{Bellucci2009}. Thus, the PAHs and HACs in Titan's atmosphere could be slightly different from those in Saturn's atmosphere.

Our study of \SI{3.4}{\um} band observed by solar occultation at \SI{145}{km} of altitude suggests the presence of aromatic molecules like PAHs and more complex like HACs at low altitudes. Some insights can be gained by examining the results from laboratory experiments. \cite{Dartois2004,Dartois2005} synthesized HACs (called {a-C:H} in these references) by ultraviolet photolysis of methane \citep{Dartois2004} and other hydrocarbons and nitrogen compounds to form {a-C:H:N} compounds \citep{Dartois2005} by UV radiation with wavelengths shorter than 120 nm in the EUV domain. Then, they made spectral Infra-Red (IR) analysis of the residues. The aim of these previous studies was to explain the spectral features observed in the ISM. Due to the presence of methane, others hydrocarbons and nitrogen compounds as well as the ultraviolet radiation in EUV domain going down to \SI{600}{km} \citep{Yoon2014}, the same HACs production process as in the ISM could occur in the Titan's upper atmosphere.

But this explanation is not completely satisfactory, mainly because we should find absorptions in the wavelength range \SIrange{3.38}{3.48}{\um} in high altitude occultation spectra which is not the case \citep{Courtin2015}. In this latter reference, the authors do not observe strong features (in extinction) in this wavelength range above \SI{500}{km} but rather below \SI{480}{km} \citep[Fig. 2]{Courtin2015}. The authors computed the \SI{3.33}{\um} / \SI{3.38}{\um} extinction coefficient ratio, the wavelength at \SI{3.33}{\um} being characteristic of aromatic \ce{C-H} stretching and the wavelength at \SI{3.38}{\um} characteristic of aliphatic \ce{C-H} stretching. Thus, this ratio is function of the ratio between the aromatic and aliphatic components in the haze particles.
This ratio is about 3 at \SI{700}{km} and decreases to about 0.5 below \SI{300}{km}. Then the conclusion of \cite{Courtin2015} is a growth of molecules from PAHs to more complex organics by particle-aging and coating process when the altitude decreases. So, even if the explanation in \cite{Dartois2004,Dartois2005} about the HACs production in the ISM is not completely relevant for Titan's upper atmosphere, it is possible to have HACs at low altitude, by particle-aging and coating process. This process was also studied in laboratory by ultraviolet irradiation of analogs of Titan's thermosphere aerosols \citep{Carrasco2018,Couturier-Tamburelli2018}.
\cite{Carrasco2018} analyzed the absorption peaks of residues and their time evolution. They found shifts and modifications of vibrational signatures reflecting aerosols transformation during the irradiation. The observed spectra after \SI{24}{h} of irradiation appear to tend toward Titan's spectra observed in VIMS spectra at altitude \SI{200}{km}. The conclusion of \cite{Carrasco2018} is that a particle aging process by ultraviolet irradiation occurs in Titan's atmosphere from aerosol embryos in thermosphere to more complex haze particles in low altitudes according the following process: (1) aerosol embryos generation, (2) sedimentation and (3) chemical evolution by UV irradiation.

Other studies were conducted \citep{Gudipati2013,Yoon2014}. In \cite{Yoon2014}, the authors experimentally studied the role of benzene photolysis in the PAHs and more complex particles (\emph{tholins}) production. They stipulated that UV radiation with wavelengths longer than \SI{130}{nm} (FUV domain) leads to the photodissociation of benzene. Likewise, in the FUV domain the photons could reach low altitudes bringing a benzene photolysis in the lower atmosphere. Then this photolysis would lead to the production of PAHs and more complex particles.
In the same way, \cite{Gudipati2013,Couturier-Tamburelli2014} proceeded to photochemical experiments with \ce{C4N2} ice and photons in FUV domain, reproducing the environment of Titan's lower atmosphere and highlighting the fact that at low altitudes (below \SI{200}{km}) the UV flux at \SI{350}{nm} is comparable to the upper radiation field at shorter wavelengths. Their conclusion is that photoabsorption by haze particles in this FUV domain would trigger a rich solid-state chemistry at low altitudes. In a similar way photolysis of \ce{HC5N} led to residues containing aromatic signatures around \SI{3.3}{\um} and \ce{C-H} stretching around \SI{3.4}{\um} \citep{Couturier-Tamburelli2015}. The authors concluded that this photolysis drives the formation of more and more complex polymers which are precursors of haze particles. These results about the PAHs and HACs production at low altitudes are in agreement with our spectral analysis of the \SI{3.4}{\um} band for the occultation at \SI{145}{km} of altitude.

Finally, NASA has recently announced the selection of the \emph{Dragonfly} mission \citep{Turtle2019} as part of its \emph{New Frontiers} program. This revolutionary quacopter is planned to explore Titan's surface, in the region of Shangri-La dune fields and Selk impact crater, during the mid-2030s. Among onboard instruments, the \emph{Dragonfly Camera Suite} will allow PAH detection via fluorescence, excited by a controlled UV illumination \citep{Lorenz2018}.

\section{Conclusion}
\label{concl}

To conclude, the present study was to explain the strong absorption around \SI{3.4}{\um} observed in the occultation spectra by VIMS at low altitudes typically lower than \SI{450}{km}. We opted for the altitude \SI{145}{km} which offers a good compromise: at lower altitudes, the stronger absorptions damp spectral features and at higher altitudes, the absorptions are too weak to be extracted from the signal. As a first step in our radiative transfer modeling, we have included 9 molecules following \cite{Maltagliati2015}: \ce{CH4}, \ce{CH3D}, \ce{CO}, \ce{C2H2}, \ce{C2H4}, \ce{C2H6}, \ce{H2O}, \ce{C6H6} and \ce{HCN}. This composition leads to a poor reproduction of the observed transmittance curve. The disagreement was reduced by including \ce{C2H6} spectroscopic data as the data coming from \cite{Pine1982} or the more exhaustive data, but empirical pseudo-line, coming from \cite{Harrison2010}. This way, the agreement has been improved, but the simulated transmittance remained above the observed one. The inclusion of propane improved the result but there was still a lack of efficient absorbers around \SI{3.4}{\um}. Knowing that spectral signatures around \SI{3.4}{\um} are present in the ISM and identified as PAHs or HACs, we have considered these complex hydrocarbon compounds in our model as possible absorbers in the Titan's atmosphere at about \SI{145}{km} of altitude. The fit of abundance of PAHs and HACs, taking into account some uncertainty by means of a correction factor, allowed a rather satisfactory modelization of the observed transmittance at \SI{145}{km} of altitude. Thus, the model suggests the presence of complex hydrocarbon compounds and precursors of haze particles at low altitudes. This result is also consistent with several laboratory studies showing a complexification of hydrogenated molecules from small hydrocarbons to more complex by UV irradiations \citep{Yoon2014,Couturier-Tamburelli2015,Couturier-Tamburelli2018,Carrasco2018}.


\clearpage

\section*{Acknowledgment}

We thank Laurence Regalia, Michael Rey, Bruno B\'{e}zard, Walter Lafferty, Christophe Sotin and Jean Vander Auvera for scientific discussion. We are grateful to Christiaan Boersma for his kind technical help in using the NASA Ames PAH database. Finally, we warmly thank Vladimir Krasnopolsky for providing us with vertical profiles of organics.

\section*{Appendix A. Spectral data references}

\begin{table}[!ht]
    \footnotesize
    \caption{
        Summary of spectral line lists used in this study.
    }
    \label{tab:lines_lists}
    \begin{tabular}{l l}
    \toprule
    Molecules & Lines source\\
    \midrule
    \ce{CH4}   & Theoretical Reims-Tomsk Spectral database $^a$ \\
    \ce{CH3D}  & Theoretical Reims-Tomsk Spectral database $^a$ \\
    \ce{CO}    & HITRAN $^b$, GEISA $^c$ \\
    \ce{C2H2}  & HITRAN $^b$, GEISA $^c$ \\
    \ce{C2H4}  & Theoretical Reims-Tomsk Spectral database $^a$ \\
    \ce{C2H6}  & HITRAN $^b$, PL82 $^d$, H10 $^e$ \\
    \ce{H2O}   & HITRAN $^b$, GEISA $^c$ \\
    \ce{C6H6}  & Ab-initio \texttt{MP2/6-211**} \\
    \ce{HCN}   & HITRAN $^b$, GEISA $^c$ \\
    \ce{C3H8}  & H10-P $^f$ \\
    \ce{C4H10} & NIST $^g$ \\
    \ce{PAHs}  & NASA AMES PAHs IR Spectroscopic database $^h$ \\
    \ce{HACs}  & \cite{Dartois2007} \\
    \hline
    \bottomrule
    \end{tabular}
    \parbox{.95\linewidth}{\scriptsize
        $^a$ \cite{Rey2016} - \href{http://theorets.tsu.ru}{theorets.tsu.ru}\\
        $^b$ \cite{Rothman2013} - \href{http://hitran.org}{hitran.org}\\
        $^c$ \cite{Jacquinet-Husson2008} - \href{http://www.pole-ether.fr/geisa}{pole-ether.fr/geisa}\\
        $^d$ \cite{Pine1982} - Available amount request.\\
        $^e$ \cite{Harrison2010} - \href{http://mark4sun.jpl.nasa.gov/pseudo.html}{mark4sun.jpl.nasa.gov/pseudo.html}\\
        $^f$ \cite{Harrison2010a} - \href{http://mark4sun.jpl.nasa.gov/pseudo.html}{mark4sun.jpl.nasa.gov/pseudo.html}\\
        $^g$ \href{http://webbook.nist.gov}{webbook.nist.gov}\\
        $^h$ \href{http://www.astrochem.org/pahdb}{astrochem.org/pahdb}
    }
\end{table}

\section*{Appendix B. List of VIMS cubes}

The list of the VIMS cubes used in the T10E, T53E, T78E and T78I occultations are provided in 4 CSV files.
They contain, the cube unique ID (in \href{https://tools.pds-rings.seti.org/opus/}{OPUS} format), its folder location on the PDS (add \href{https://pds-rings.seti.org/holdings/volumes/COVIMS_0xxx/}{pds-rings.seti.org/holdings/volumes/COVIMS\_0xxx/} prefix to get the URL), the cube acquisition start time and the observation sequence name.

\bibliography{biblio}

\end{document}